\documentclass[10pt,a4paper,twocolumn,superscriptaddress,aps,raggedbottom]{revtex4}

\usepackage[a4paper,
            top=2.54cm,
            bottom=2.54cm,
            left=1.5cm,
            right=1.5cm]{geometry}

\usepackage{xcolor}
\usepackage{amsmath}
\usepackage{graphicx}
\usepackage{hyperref}

\makeatletter
\renewcommand{\figurename}{Fig.}
\renewcommand{\fnum@figure}{\textbf{\figurename~\thefigure}}
\makeatother

\begin{document}

\title{Laser electro-optic frequency comb in lithium niobate nanophotonics}

\author{Benjamin~K.~Gutierrez}
\altaffiliation{These authors contributed equally.}
\affiliation{Department of Applied Physics and Materials Science, California Institute of Technology, Pasadena, California 91125, USA.}

\author{Yingchu~Xu}
\altaffiliation{These authors contributed equally.}
\affiliation{Department of Electrical Engineering, California Institute of Technology, Pasadena, California 91125, USA.}

\author{Nicolas~Englebert}
\altaffiliation{These authors contributed equally.}
\affiliation{Department of Electrical Engineering, California Institute of Technology, Pasadena, California 91125, USA.}

\author{Rithvik~Ramesh}
\affiliation{Department of Electrical Engineering, California Institute of Technology, Pasadena, California 91125, USA.}

\author{Maximilian~Shen}
\affiliation{Department of Electrical Engineering, California Institute of Technology, Pasadena, California 91125, USA.}

\author{Deven~Tseng}
\affiliation{Department of Electrical and Computer Engineering, University of California, Santa Barbara, Santa Barbara, California 93106, USA.}

\author{Adelynn~Tang}
\affiliation{The Division of Physics, Mathematics and Astronomy, California Institute of Technology, Pasadena, California 91125, USA.}

\author{Ryoto~Sekine}
\affiliation{Department of Electrical Engineering, California Institute of Technology, Pasadena, California 91125, USA.}
\affiliation{PINC Technologies Inc., Pasadena, California 91125, USA.}

\author{Mahmood~Bagheri}
\affiliation{Jet Propulsion Laboratory, California Institute of Technology, Pasadena, CA 91009, USA.}

\author{Auro~M.~Perego}
\affiliation{Aston Institute of Photonic Technologies, Aston University, Birmingham B4 7ET, UK.}

\author{Alireza~Marandi}
\email{marandi@caltech.edu}
\affiliation{Department of Applied Physics and Materials Science, California Institute of Technology, Pasadena, California 91125, USA.}
\affiliation{Department of Electrical Engineering, California Institute of Technology, Pasadena, California 91125, USA.}


\begin{abstract} %
\noindent Optical frequency combs have revolutionized precision science and technology, yet their nanophotonic implementations have failed to simultaneously achieve high efficiency, power, and coherence. Optically driven microcombs provide broad and stable spectra but low usable power, whereas active comb generators, including mode-locked lasers, can be efficient yet offer less control over coherence. We introduce the laser electro-optic (LEO) frequency comb, a comb-generation mechanism in which coherent continuous-wave injection drives a phase-modulated laser cavity above threshold into a distinct operating regime. Unlike other coherently driven integrated comb sources, the LEO comb generates comb powers that exceed the injected continuous-wave power by an order of magnitude. We realize the LEO comb in a hybrid lithium niobate/III-V nanophotonic circuit and demonstrate milliwatt-level power per comb line, 1.76-ps pulses, a 4.7-nm background-free spectrum, and linewidths as narrow as 19.6 kHz. By unifying high efficiency, power, and coherence, this architecture establishes a definitive route to chip-scale frequency comb sources that deliver on the promise of scalable, high-performance coherent optical technologies.
\end{abstract}

\maketitle
\noindent Optical frequency combs (OFCs) have become indispensable tools across a wide range of applications, including optical frequency metrology, spectroscopy, LiDAR, and optical frequency division~\cite{udem_optical_2002,coddington_coherent_2008,coddington_rapid_2009,fortier_generation_2011}. 
Their broad utility has motivated substantial efforts towards their miniaturization, leading to the development of many integrated sources~\cite{suh_microresonator_2016,trocha_ultrafast_2018,kippenberg_dissipative_2018,faist_quantum_2016,guo_ultrafast_2023,davenport_integrated_2018,hermans_high-pulse-energy_2021,yu_integrated_2022, chang_integrated_2022}. Yet integrated OFCs remain constrained by a trade-off among efficiency, output power, and coherence. Although coherently driven passive comb generators can provide broad and stable spectra, they have typically exhibited pump-to-comb conversion efficiencies below a few percent~\cite{yang_efficient_2024}, and despite recent improvements in performance~\cite{xue_microresonator_2017,hu_high-efficiency_2022,boggio_efficient_2022,helgason_surpassing_2023}, they still fall short of the wall-plug efficiencies of diode lasers. Active comb generators can achieve high electrical-to-optical efficiency, but integrated implementations, often based on semiconductor optical gain, generally offer less control over coherence and absolute optical frequencies. \\\indent
Among the various integrated comb-generation approaches, electro-optic (EO) frequency combs offer a particularly attractive route, as they generate combs by modulating a continuous-wave (CW) laser with readily available electro-optic modulators (EOMs). Early work showed that intracavity phase modulation could increase the modulation strength through resonant enhancement~\cite{gordon_fabry-perot_1963}. These resonant EOMs were first used to demonstrate high repetition rate pulse generators\,\cite{kobayashi_highrepetitionrate_1972} and subsequently resonant electro-optic (REO) frequency comb generators\,\cite{kourogi_wide-span_1993} in cavities resonant either at the CW pump frequency\,\cite{ kovacich_short-pulse_2000, xiao_toward_2008} or at its half-harmonic through the quadratic nonlinear interactions\,\cite{diddams_broadband_1999}. As EOM technology matured, single-pass EO combs emerged~\cite{Fujiwara:03}, enabling electronic control of the comb line spacing through the radio-frequency (RF) modulation~\cite{Wu:10}, broadening the appeal of EO combs for diverse applications\,\cite{parriaux_electro-optic_2020}.\\\indent 
More recently, REO combs have been realized in integrated photonic platforms~\cite{zhang_broadband_2019, zhang_ultrabroadband_2025}, thanks to the availability of thin-film materials with a strong electro-optic effect, namely thin-film lithium niobate (TFLN)\,\cite{zhu_integrated_2021} and thin-film lithium
tantalate\,\cite{wang_lithium_2024}. Yet, akin to Kerr microcombs\,\cite{coen_universal_2013}, the high-Q realization of REO combs for enhancing the spectral coverage, substantially limits power extraction, resulting in optical conversion efficiencies less than 1\%\,\cite{zhang_broadband_2019}.
To address this limitation, several schemes have been developed to increase the optical energy conversion efficiency of EO combs. For example, time lensing enhances this efficiency by shortening the pulses and increasing the output peak power, achieving optical conversion efficiencies as high as 25\%~\cite{yu_integrated_2022}. Moreover, coupled-resonator EO combs have pushed optical conversion efficiencies upwards of 32\%, enabled by efficient energy flow into the comb cavity via an auxiliary resonator~\cite{hu_high-efficiency_2022}. Nevertheless, these passive EO architectures generate usable comb powers that are strictly lower than the input injection power, as they result from dispersing the input CW frequency across multiple comb lines.\\\indent
\begin{figure*}[t]  
    \centering
    \hspace{-5mm}
    \includegraphics[width=183mm]{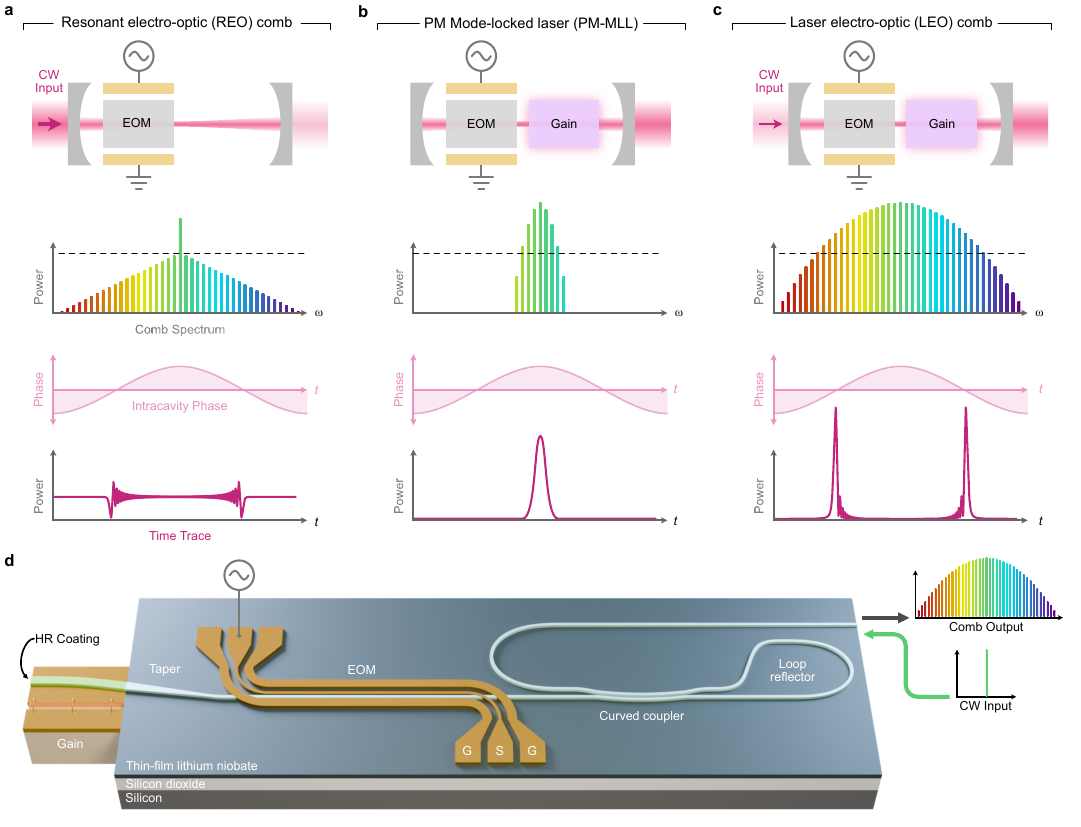}
    \caption{\textbf{LEO comb concept.} 
    \textbf{a}, A REO comb, which includes an optical resonator with an intracavity phase modulator subject to a CW injection, along with its generated comb spectrum, intracavity phase, and output time trace.
    \textbf{b}, A PM-MLL, which includes a laser cavity with an intracavity phase modulator without any external optical injection, along with its generated comb spectrum, intracavity phase, and output time trace.
    \textbf{c}, The LEO comb, which includes a laser cavity with an intracavity phase modulator and a CW injection, along with its generated comb spectrum, intracavity phase, and output time trace. Both panels a and c consider the CW injection at the center of a cavity resonance, i.e., zero detuning between the injection and cavity.
    \textbf{d}, Schematic of the integrated LEO comb implemented on a TFLN  nanophotonic circuit and an electrically-pumped semiconductor gain element. HR, highly reflective; GSG, ground signal ground.} 
    \label{fig:Panel1}
\end{figure*}
Alternatively, efficient comb generation can be achieved in active cavities through a mechanism fundamentally different from that of passive EO combs, arising from the interplay between intracavity phase modulation and intrinsic laser dynamics~\cite{siegman_lasers_1986,keller_ultrafast_2021}. In integrated photonics, recent demonstrations have explored two main regimes: (i) quantum walk combs, which arise when the gain recovery time is faster than the cavity round-trip time and ultra-short pulse formation is unlikely~\cite{heckelmann_quantum_2023,cargioli_quantum_2025,letsou_high-power_2025,marzban_quantum_2026}, and (ii) mode-locked lasers (MLLs), in which the gain recovery time is slower than the cavity round-trip time enabling the formation of well-defined pulses that can approach the transform limit when dispersion and filtering are properly managed~\cite{guo_ultrafast_2023,ling_electrically_2024,zhang_broadband_nodate}. However, in contrast to EO combs, these demonstrations do not include an external CW injection, which could otherwise enable control over comb coherence, phase noise, and absolute frequencies. Furthermore, while they may provide higher wall-plug efficiencies through direct electrical-to-optical comb generation, they exhibit relatively narrow spectra with respect to EO combs. These limitations highlight the need for a comb source that combines the coherence and spectral bandwidth of passive EO combs with the efficiency and output power of an active laser cavity. \\\indent 
In this work, we realize this combination in a hybrid TFLN/III-V nanophotonic platform by implementing a phase-modulated laser cavity subject to optical injection from a CW laser, enabling the formation of an electro-optic frequency comb that exists above the lasing threshold, due to the low saturation power of the semiconductor gain. In this configuration, lasing action allows the comb lines to grow much larger than the input optical power, generating a background-free spectrum, spanning up to 15 nm, with more than 33 times greater power than the CW injection, a 100-fold improvement over state-of-the-art REO comb architectures~\cite{hu_high-efficiency_2022}, and resulting in milliwatt-level power per comb line. Furthermore, we attain 1.76-ps pulses directly from the laser cavity, surpassing what can be achieved by a phase-modulated MLL (PM-MLL) in the same configuration~\cite{guo_ultrafast_2023,haus_mode-locking_2000}. Finally, the low-noise CW laser injection reduces the free-running semiconductor linewidth to comb lines as narrow as 19.6~kHz, yielding a highly coherent comb.

\subsection*{Concept and Theory}
\begin{figure*}[t]  
    \centering
    \hspace{-5mm}
    \includegraphics[width=183mm]{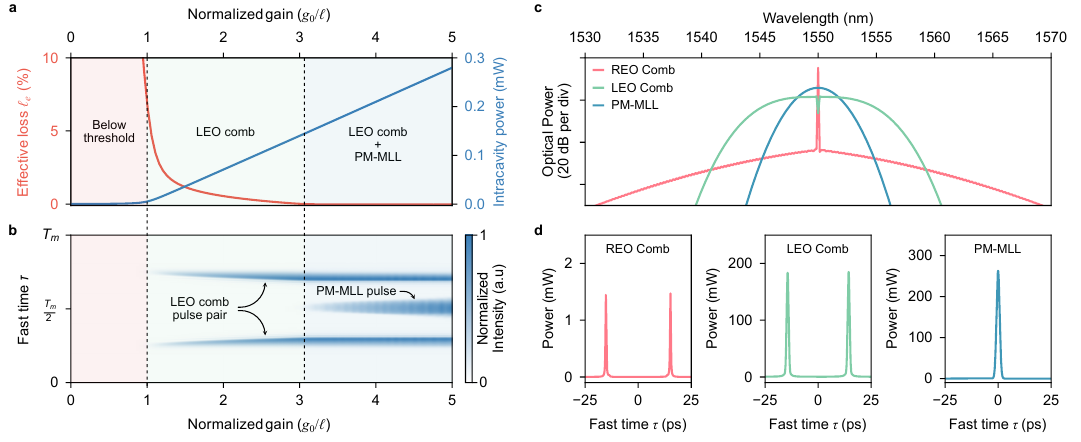}
    \caption{
    \textcolor{black}{
        \textbf{LEO comb simulations and comparisons.}    
        \textbf{a}, Average intracavity power (blue line) and percent effective loss (red line) plotted against normalized gain.
        \textbf{b}, 2D colormap of the intracavity power for the fast time plotted versus normalized gain. The three shaded regions indicate the below threshold regime (red shade), the LEO comb regime (green shade), and phase-modulated mode-locking with LEO comb regime (blue shade), in both panels a and b. 
        \textbf{c}, Numerical simulations showing the spectral power density advantage of LEO comb compared with REO comb and PM-MLL, even with their further amplification. Spectra of the post-amplified REO comb, LEO comb, and PM-MLL with the same M, $\delta_0$, $\omega_m$, and $T_R$.
        \textbf{d}, The corresponding pulses for the spectra in panel c. 
    }
    }
    \label{fig:Panel2}
\end{figure*}

\noindent The concept of LEO comb and how it differs from REO comb and PM-MLL is illustrated in Fig.\,\ref{fig:Panel1}. The LEO comb consists of an optically injected resonator containing intracavity phase modulation and optical gain, which appears as the combination of a REO comb (Fig.\,\ref{fig:Panel1}a) and a PM-MLL (Fig.\,\ref{fig:Panel1}b) in terms of its building blocks. However, its operating principle is fundamentally distinct from them.\\\indent 
In contrast to a passive REO comb, in the presence of large gain, when the modulation frequency matches the free spectral range of the resonator, the generated sidebands resonate in the laser cavity, and a broadband comb can oscillate in the cavity above its lasing threshold, leading to powers far exceeding the injected power. The LEO comb and REO comb chirped pulses form at the inflection points of the intracavity phase modulation, as illustrated in the time plots of Fig.\,\ref{fig:Panel1}a,c. However, the large intracavity gain of the LEO comb yields pulse energies and peak powers that are orders of magnitude higher than those of the REO comb, owing to above threshold operation.  \\\indent
In contrast to a PM-MLL, the LEO comb leverages an external CW injection to saturate the gain medium while maintaining a non-zero effective loss. This unique physical state prevents the transition to free-running lasing, ensuring that the external injection remains the primary determinant of the comb’s phase and temporal dynamics. Crucially, in a PM-MLL, the pulses form at the turning points of phase modulation (Fig.\,\ref{fig:Panel1}b)\,\cite{Ramesh:25}. This timing difference represents the fundamental distinction between the pulse formation in the LEO comb versus the PM-MLL, and signifies why the LEO comb cannot be considered an injection-locked variant of the PM-MLL. \\\indent
Figure \ref{fig:Panel1}d shows the schematic of the LEO comb source, in an integrated Fabry-Perot laser configuration consisting of a TFLN nanophotonic circuit and an electrically pumped semiconductor gain element. The laser cavity is formed between a gain chip with a highly reflective back facet and a TFLN external cavity, which implements an intracavity EOM and a Sagnac loop reflector. The CW injection is provided through the output port of the laser.\\\indent
To theoretically study LEO comb formation, we model our system with a generalized driven Haus master equation~\cite{haus_mode-locking_2000, perego_coherent_2020}, 
\begin{align}
\hspace{-2mm} T_R\,\frac{\partial A}{\partial T}= &
\Bigg[
-\ell+\frac{g_0}{1+\langle |A|^2 \rangle/P_{\mathrm{sat}}}
\Big(1 + \frac{1}{\Omega^2_g}\frac{\partial^2}{\partial \tau^2}\Big)
 \label{eq1} \\ \notag
&\hspace{-15mm}-i[\delta_0-M\cos(\omega_m \tau)]
-i\dfrac{\beta_2L}{2}\,\frac{\partial^2}{\partial \tau^2}+i\gamma L |A|^2\Bigg] A 
+ \sqrt{\theta}E_\text{in}
\end{align}
where $T$ is a slow time describing the evolution of the intracavity field, $A(T,\tau)$, over consecutive round trips, $T_R$ is the round-trip time, and $\tau$ is a fast time used to describe the intracavity dynamics. $\ell$ is the total round-trip loss. The gain medium is modeled by an unsaturated gain term, $g_0$, with saturation power, $P_\text{sat}$, and gain bandwidth $\Omega_g$. Here, 
$\langle |A|^2 \rangle=\int_0^{T_R}|A|^2/T_R\,\text{d}
\tau$ denotes the average intracavity power over one round-trip, $\delta_0$ denotes the injection frequency detuning and $L$ denotes the round-trip cavity length. The intracavity phase modulation has frequency $\omega_m$ and modulation amplitude $M$. $\beta_2$ and $\gamma$ correspond to the averaged group velocity dispersion and third-order effective nonlinear coefficients, respectively. Finally, $\theta$ is the coupler transmission coefficient, and $E_\text{in} = \sqrt{P_\text{in}}e^{i\omega_\text{in}\tau}$, where $P_\text{in}$ is the power of the optical injection and $\omega_\text{in}$ its frequency.\\\indent
Figure\,\ref{fig:Panel2}a shows the numerical simulation results of different operating regimes of the LEO comb, which include both the average intracavity power as well as the effective loss, defined as $\ell_e = \ell-g$, where $g$ is the saturated gain, as a function of the normalized gain, $g_0/\ell$. The intracavity power along the fast time, $\tau$, for different values of normalized gain is shown in Fig. \ref{fig:Panel2}b. Three different regions can be clearly identified. When gain is less than total round-trip loss ($g_0 < \ell$), the laser operates below its oscillation threshold. Beginning at $g_0 / \ell = 1$, the average power increases linearly with gain, similar to the typical behavior of above-threshold lasers. However, the effective loss does not clamp to zero, as it would in the absence of injection, owing to the additional saturation caused by the injection. In this second region, where $g_0 > \ell$ but $\ell_e \neq 0$, the linear increase in intracavity power is attributed to lasing at the frequency of the injection. As evident in Fig. \ref{fig:Panel2}b, this second region produces a pair of short pulses located at the inflection point of the phase modulation profile, for on-resonance injection, as shown previously in Fig. \ref{fig:Panel1}c. We refer to this region as the LEO comb regime, whose phase-locked frequency comb spectrum is shown in Fig. \ref{fig:Panel2}c for $g_0 /\ell =$ 1.35. As the gain is increased further, the effective loss approaches zero as the saturation caused by the injection becomes negligible, leading to the third region where $g_0 > \ell$ and $\ell_e = 0$. As a result, PM-MLL pulses appear at the turning point of the phase modulation signal, as shown in Fig. \ref{fig:Panel2}b (see also Fig. \ref{fig:Panel1}b). The third region represents a combination of a LEO comb with a PM-MLL. The impacts of the injection detuning and gain saturation power on the LEO comb regime are discussed in the SI, sections II-III.\\\indent
To demonstrate the substantial pump-to-comb efficiency advantage of LEO comb over REO comb and its superior spectral coverage over PM-MLL, we perform side-by-side numerical simulations using the same components for these three cases, as shown in Fig. \ref{fig:Panel2}c,d. For the REO comb, we set $g_0 = 0$ in equation~\eqref{eq1}. However, for fair comparison, we then amplify the REO comb output by the same $g_0$ and $P_{sat}$ as in the LEO comb, keeping all other parameters identical except the total round-trip loss, which we set to 1\% and enforce critical coupling to achieve a high Q-factor and broadband operation~\cite{buscaino_design_2020}. The resulting post-amplified REO comb spectrum is shown in Fig. \ref{fig:Panel2}c, alongside the LEO comb spectrum. Additionally, we plot the PM-MLL spectrum using the same parameters as the LEO comb but in the absence of injection with $P_{in} =$ 0. The three corresponding temporal traces are given in Fig. \ref{fig:Panel2}d, where we filter the CW background of the REO comb to elucidate its pulse shape.
With the same constituent components, the LEO comb generates peak powers more than two orders of magnitude higher than the passive REO comb, owing to lasing dynamics that enable high-power buildup and efficient power extraction from the resonator. Moreover, it generates a broader comb spectrum and shorter pulse lengths than the PM-MLL, resulting from its fundamentally different comb formation mechanism.
\begin{figure*}[tp]  
    \centering
    \hspace{-5mm}
    \includegraphics[width=183mm]{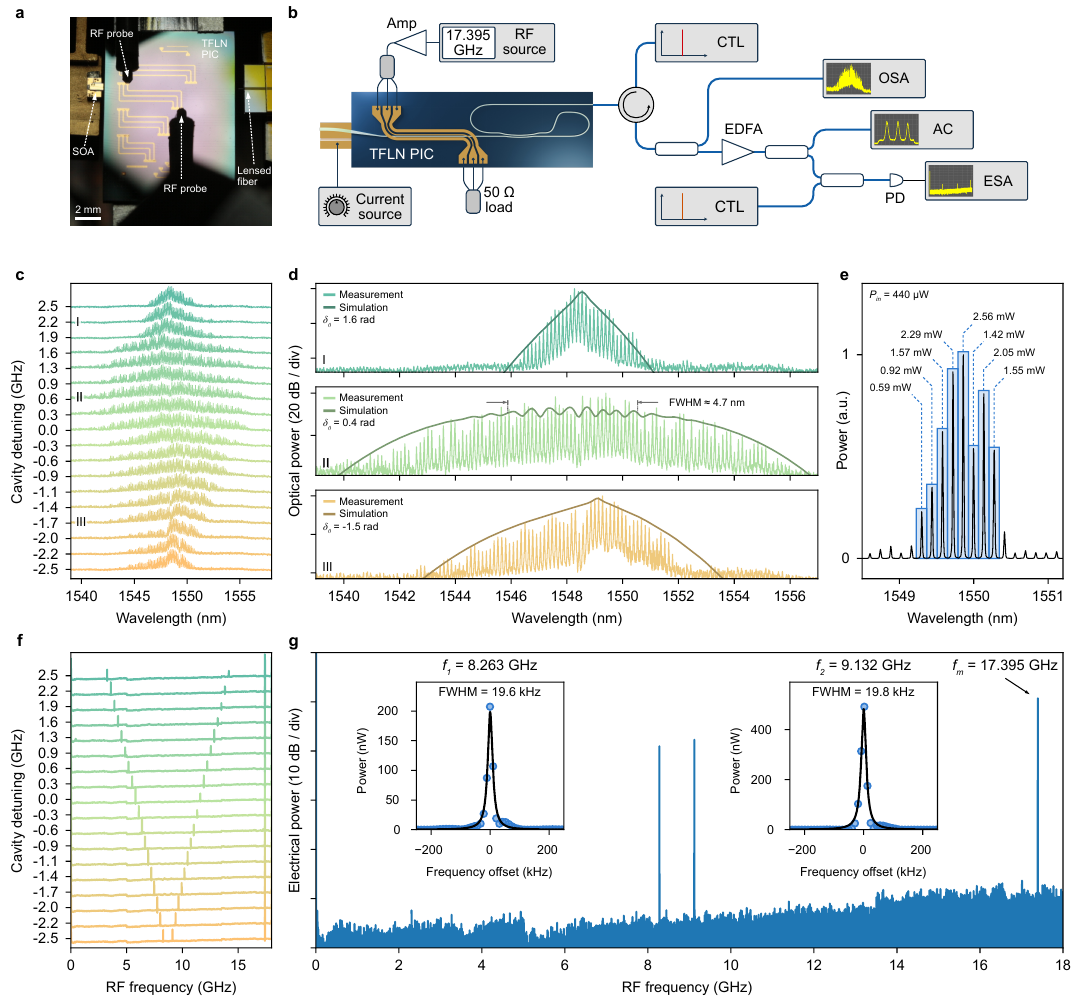}
    \caption{
    \textcolor{black}{
        \textbf{LEO comb experiments.}    
        \textbf{a},\,Image of device in the test stand. PIC, photonic integrated circuit.
        \textbf{b},\,Schematic of the experimental setup. Amp., RF Amplifier; EDFA, erbium-doped fiber amplifier; PD, photodiode. 
        \textbf{c},\,Comb spectra as the injection frequency was tuned across 5 GHz near 1548 nm.
        \textbf{d},\,Theoretical spectral envelopes at $\delta_0$ = 1.6 rad, 0.4 rad, and -1.5 rad (dark lines), shown together with the corresponding experimental traces labeled I, II, and III in panel c, respectively.
        \textbf{e},\,Normalized linear power spectrum showing integrated power per comb line for comb lines over 0.5 mW. The shaded boxes indicate the integration region for each comb line.
        \textbf{f},\,Heterodyne beat notes as the injection frequency was tuned across 5 GHz near 1548 nm. The colors correspond to data taken at different injection frequencies, simultaneously acquired during the sweep of panel c.
        \textbf{g},\,Heterodyne beat note measurement, showing the modulation frequency at $f_m$ = 17.395 GHz and the frequencies of the two resulting beat notes. The insets show two beat notes, with Lorentzian fits corresponding to FWHM values of 19.6\,kHz and 19.8\,kHz, respectively.
    }    
    }
    \label{fig:Panel3}
\end{figure*}

\subsection*{Spectral characteristics}
\noindent Next, we experimentally demonstrate LEO comb formation. We use a nanophotonic TFLN chip edge coupled to an SOA, followed by a lensed tip fiber at its output, as depicted in Fig. \ref{fig:Panel3}a. The TFLN chip implements an intracavity phase modulator consisting of a coplanar waveguide (CPW) transmission line designed to achieve a 50 $\Omega$ impedance. The schematic of the experimental setup is shown in Fig. \ref{fig:Panel3}b, which includes a continuously tunable laser (CTL) for optical injection into the integrated laser cavity, an optical spectrum analyzer (OSA), an intensity autocorrelator (AC), and an electrical spectrum analyzer (ESA), which are used for the simultaneous acquisition of the laser spectrum, autocorrelation time trace, and heterodyne beat notes.\\\indent 
We start by applying a 17.395 GHz RF signal to the intracavity EOM corresponding to the second harmonic of the cavity FSR and scanning the optical injection frequency across a cavity resonance. The resulting spectra are shown in Fig.\,\ref{fig:Panel3}c, which are obtained when the laser is operating in the LEO comb regime, as marked in Fig.\,\ref{fig:Panel2}a,b. We observe a significant change in the optical spectrum as the injection frequency is tuned. Specifically, at larger detuning values, the spectral bandwidth is relatively narrow, as indicated by (I), resulting in the concentration of the comb power in fewer comb lines. As the optical injection is tuned near the cavity resonance ($\delta_0 \sim 0$), a broad spectral bandwidth is generated, as shown by (II). The LEO comb evolution with respect to the injection frequency is similar to that of a REO comb, where the broadest spectra occur when the injection frequency matches a cavity resonance~\cite{zhang_broadband_2019}.\\\indent
To verify the spectral behavior of the LEO comb, we include the simulated spectra, obtained by numerically integrating equation\,\eqref{eq1}, alongside the measured spectra for the three selected operating points, shown in Fig. \ref{fig:Panel3}d. The three operating points correspond to the spectra marked by I, II, and III in Fig. \ref{fig:Panel3}c. Changing the optical injection frequency results in simulated spectra that match well with the measurements at those detuning values. At a detuning of 600 MHz, a broad comb emerges, whose comb lines span over 15 nm with a 4.7 nm FWHM. In the absence of optical driving, equation\,\eqref{eq1} predicts a spectral bandwidth of around 1.7 nm for a PM-MLL with identical simulation parameters. The LEO comb produces background-free spectra with a nearly threefold broader bandwidth when compared to a PM-MLL in the same setup.\\\indent
\begin{figure*}[t] 
    \centering
    \hspace{-5mm}
    \includegraphics[width=183mm]{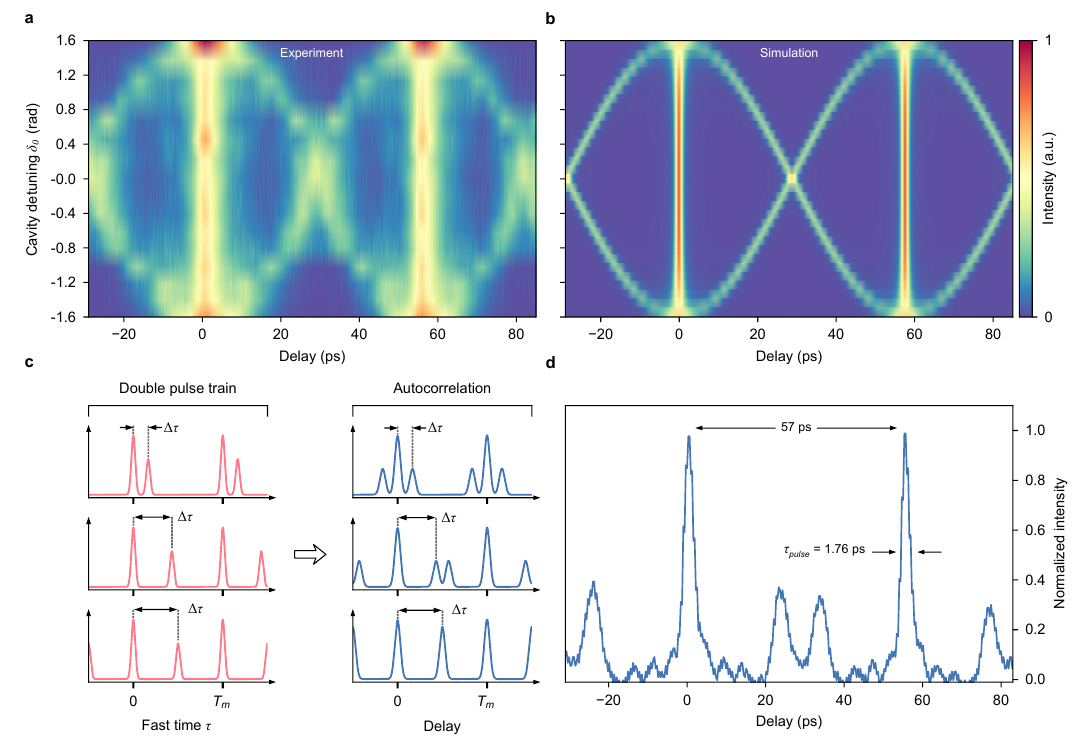}
    \caption{
    \textcolor{black}{
        \textbf{LEO comb pulse dynamics measurements and simulations.}
        Measured \textbf{a}, and simulated \textbf{b}, autocorrelation traces versus cavity detuning.
       \textbf{c}, Illustration of a pulse pair train (left in red) and the resulting autocorrelation function for that trace (right in blue). From top to bottom, the pulse pair spacing, $\Delta\tau$, increases, resulting in the side features of the autocorrelation trace spreading by the same amount.
        \textbf{d}, Autocorrelation trace of the LEO comb. Assuming a Gaussian pulse shape, the measured autocorrelation width corresponds to a pulse FWHM of 1.76\,ps. With $P_\text{in}=151\,$µW of optical driving power, the LEO comb produces 2.33 mW of average power corresponding to 35.8 mW peak power, which is 237 times greater than the input CW optical power. 
    }
    }
    \label{fig:Panel4}
\end{figure*}
We further demonstrate large power extraction from the gain medium. Leveraging lasing action, the LEO comb builds up significant power across the comb, directly generating comb lines with milliwatt-level power from sub-milliwatt-level injection powers. To achieve this, we increase the current of the SOA and tune the injection to achieve a narrower spectrum with high power per comb line, as shown in Fig. \ref{fig:Panel3}e. The laser produces 14.67 mW of on-chip power, 33.3 times the input CW injection power ($P_{in}$ = 440 µW). Furthermore, we observe 8 comb lines whose integrated power exceeds the input power and 6 adjacent comb lines with over 1 mW each. Such a high efficiency distinguishes the LEO comb from other integrated OFC sources\,\cite{chang_integrated_2022}.\\\indent 
Figure \ref{fig:Panel3}f shows the heterodyne beat notes acquired during the injection frequency scan of Fig. \ref{fig:Panel3}c, and confirms the comb formation along the entire tuning range. As the injection is tuned, the comb lines follow this injection frequency, causing the beat notes to move by the same value. This highlights the comb's robustness against injection detuning and enables fine control over the comb line frequencies.\\\indent
A crucial characteristic of OFCs is their optical linewidth. When compared to coherently driven passive comb generators, free-running semiconductor MLLs typically exhibit broader linewidths, resulting from amplitude-phase noise coupling in the gain medium~\cite{Henry_linewidth_enhancement_1982}. Here, despite the presence of a strong amplification due to being above the CW lasing threshold, we measure linewidths in the kHz regime. As shown in Fig. \ref{fig:Panel3}g, with a modulation frequency of $f_m = 17.395$\,GHz, we obtain two beat notes at $f_1 = 8.263$\,GHz and $f_2 = 9.132$\,GHz, with fitted Lorentzian FWHM linewidths of 19.6\,kHz and 19.8\,kHz, respectively, which are likely limited by the linewidth of the injection laser. This shows that while the LEO comb benefits from strong amplification, it also exhibits reduced phase noise due to its shared coherence with the injection source.
\subsection*{Temporal dynamics}
\noindent Finally, we study the pulse formation dynamics of the LEO comb and show the measured and simulated autocorrelation traces from the injection frequency scan in Fig.\,\ref{fig:Panel4}a and b, respectively. Each pulse accumulates a round-trip phase equal to the detuning, $\delta_0$. In response, the pulses settle at $\tau = \tau_0$, where the RF signal supplies adequate phase to nullify the pulses' round-trip phase (i.e., $\delta_0 + M\cos(\omega_m \tau_0) = 0$). As a result, a pulse pair is formed whose spacing is directly dependent on the injection frequency detuning. This dependence highlights the LEO comb phase-locking to the injection, consistent with the behavior of REO combs~\cite{hu_high-efficiency_2022}, despite operation above the CW lasing threshold. This pulse pair repeats itself once every modulation cycle, generating a train of pulse pairs spaced by the modulation period, $T_{m}$. Figure \ref{fig:Panel4}c illustrates how this pulse pair spacing translates to an intensity autocorrelation measurement. A double pulse generates a symmetric three-featured autocorrelation trace, whose time delay between the central peak and side features is equal to the pulse spacing, $\Delta\tau$, of the double pulse. As the pulse spacing increases, the side features of the autocorrelation trace move farther from their respective central peaks. The measured autocorrelation scan of Fig. \ref{fig:Panel4}a shows that detuning the injection across 3.2 radians (4.43 GHz) results in pulse spacings spanning zero to $T_{m}$ = 57\,ps.\\
\indent We further demonstrate the formation of a 1.76-ps pulse in the LEO comb, shown in Fig. \ref{fig:Panel4}d. We estimate its peak power to be 237 times greater than the average optical injection power. Following equation~\eqref{eq1}, this pulse length is shorter than what can be achieved from a PM-MLL in the same cavity (see Fig.\,\ref{fig:Panel2}d), which is estimated to be 2.6 ps. Thus, the LEO comb outperforms a PM-MLL in the same configuration in terms of spectral bandwidth, comb linewidth, and pulse length.
\subsection*{Discussion and outlook}
\noindent In conclusion, we demonstrate a LEO frequency comb in a hybrid TFLN/III-V integrated cavity as a highly coherent and highly efficient OFC source. We experimentally show and numerically confirm the unique dynamics of LEO comb formation, contrasting them with those of REO combs and MLLs. The demonstrated LEO comb source uniquely translates a CW injection in the microwatt regime to a coherent frequency comb with milliwatt-level power per comb line. We attain peak powers 237 times larger and comb powers 33.3 times larger than the injected CW power, reducing the required injection power and representing the highest pump-to-comb conversion among existing integrated schemes with CW injection. We observe linewidths as narrow as 19.6 kHz. We show spectral bandwidths of 4.7 nm at the 3-dB level, and pulse lengths as short as 1.76 ps, outperforming the capabilities of a PM-MLL in the same platform~\cite{guo_ultrafast_2023}. LEO comb formation further enables robustness against injection detuning, maintaining a coherent comb across 5 GHz near 1550 nm, enabling spectral tailoring, giving rise to either broadband or high power per comb line operation. \\\indent
Optimization of the input and output coupling by including state-of-the-art coupling schemes with losses as low as 1.5 dB~\cite{he_low-loss_2019, siddharth_ultrafast_2025} can lead to further enhancement of the efficiency and higher output powers. Faster modulation frequencies may be used to generate a spectrum as broad as 28.8\,nm at the 3-dB level (see SI section IV). More advanced integrated LEO comb architectures can include integrated tunable CW lasers in TFLN~\cite{li_integrated_2022,op_de_beeck_iiiv--lithium_2021} or self-injection-locking schemes instead of external injection.\\\indent
The LEO comb combines the narrow linewidth and stability of a coherent injection with the power and efficiency characteristics of a semiconductor laser, offering a path towards highly coherent and highly efficient integrated sources of OFCs. The available high peak power and power per comb line of LEO combs may enable a variety of applications from nonlinear photonics and LiDAR to communications.
\vspace{-10pt}
\renewcommand{\bibsection}{\section*{References}} 
\bibliographystyle{ieeetr}
\bibliography{LeoLib_V4_stripped}


\section*{Methods}

\noindent\textbf{Device fabrication}
\\
The LEO comb was fabricated from a 700\,nm x-cut MgO-doped thin-film lithium niobate on 4.7\,µm of SiO$_2$ with a 500\,µm silicon handle (NANOLN). Electron beam lithography is used to pattern the nominally 800\,nm wide waveguides using hydrogen silsesquioxane as the resist. Next, the TFLN is etched 350\,nm deep using Ar\textsuperscript{+} plasma, leaving a $\sim 60^\circ$ sidewall angle. Plasma-enhanced chemical vapor deposition is used to deposit 800\,nm of SiO\textsubscript{2} cladding. The RF electrodes are then patterned on top of this cladding using electron beam lithography, this time using PMMA as the resist. Electron-beam evaporation is then used to deposit 10\,nm of Cr, followed by 300\,nm of Au. Acetone is used to perform liftoff, leaving behind the RF coplanar waveguides with a 28\,µm / 8\,µm / 28\,µm (GSG) electrode widths, and an electrode gap of 4\,µm; designed to achieve a 50~$\Omega$ impedance. Finally, the waveguide facets are mechanically polished to improve facet coupling.\\

\noindent\textbf{Experimental Setup}
\\
The device is placed on a fixed stage where a gain chip is aligned to the device facet with a 6-axis nanopositioning stage, forming a Fabry-Perot cavity between the highly reflective ($>$90\%) normal facet of the C-band single-angled facet SOA (Thorlabs SAF1126C) and a loop reflector employing a curved directional coupler, which offers broadband reflection ($\sim$70\%) around the center frequency of the gain chip~\cite{morino_reduction_2014,chen_broadband_2017}. Efficient power coupling between the device and SOA was achieved by a mode-matching taper; the angle of which was designed to match the exit angle of the amplifier, considering refraction\,\cite{guo_ultrafast_2023}. The RF signal was generated by a signal generator (Rohde \& Schwarz SMA100B) and amplified by a high-power RF amplifier (Mini-Circuits ZVE-3W-183+) before being applied to the CPW phase modulator on the device. An RF power meter (Ladybug LB680A) performed an RF power measurement and provided a 50 $\Omega$ termination to the transmission line. The RF frequency was tuned to match the second harmonic of the cavity FSR, resulting in harmonic mode-locking. The laser output was coupled into a lensed fiber, which then splits, via a fiber circulator, into two paths: the injection path and the measurement path. The injection path consisted of a tunable laser (Toptica CTL 1550), which was used for optical injection into the laser cavity. The measurement path consisted of an optical spectrum analyzer (Yokogawa AQ6370D), an intensity autocorrelator (Femtochrome FR-103XL), and an electrical spectrum analyzer (Rohde \& Schwarz FSW) for the simultaneous acquisition of the laser spectrum, autocorrelation time trace, and heterodyne beat note. The signal in the measurement path was first split with a 50:50 coupler, sending one path directly to the OSA and the other path to an erbium-doped fiber amplifier (Thorlabs EDFA100P). After the fiber amplifier, a 90:10 coupler sent a majority of the amplified signal to the autocorrelator and the rest to a heterodyne detection scheme, consisting of a 50:50 coupler which combined a local oscillator from a second CW laser with the comb output, which was then passed to a fast photodiode (Thorlabs DX30AF) whose electrical signal was detected by the ESA. For the results of Figure \ref{fig:Panel3}f, the local oscillator for the heterodyne measurement consisted of a tunable CW laser (Santec TSL-510), whereas the local oscillator for the results of Figure \ref{fig:Panel3}g consisted of an ultra-low noise fiber laser (NKT Koheras ADJUSTIK X15) to ensure the linewidth measurement was not limited by the local oscillator.\\

\noindent\textbf{Injection tuning sweep}
\\
We modulated the EOM at 17.395 GHz, matching the second harmonic of the cavity FSR to further increase the LEO comb bandwidth, as suggested by Eq. (1). The SOA was supplied with 139.8 mA of current, producing 3.67 mW of on-chip power (see SI section V for coupling loss characterization). To access the LEO comb regime, the tunable laser injected 160 µW of on-chip power at 1548.23 nm. We then adiabatically tuned the optical injection frequency by 5 GHz across a cavity resonance, while the spectrum (Fig. \ref{fig:Panel3}c), heterodyne beat notes (Fig. \ref{fig:Panel3}f) and autocorrelation trace (Fig. \ref{fig:Panel4}a) were simultaneously recorded.\\

\noindent\textbf{Power per comb line}
\\
For operating at high power per comb line, we increased the SOA current to 289\,mA, producing 14.67\,mW of on-chip power. We set the on-chip injected power to 440\,µW at 1550\,nm. While maintaining LEO comb operation, we detuned the injection from the cavity resonance to a value similar to the top panel shown in Fig. \ref{fig:Panel3}d, thereby increasing the power density of the spectrum. To estimate the comb-line power, we integrated across the measured spectral trace and applied a scaling factor such that the integrated power was equal to the estimated on-chip power. The boundaries of the integration windows for each comb line were set halfway between its peak value and the peak value of the neighboring comb lines. We report the power per comb line as the sum of the power in each integration window. The power contained in the eight integration windows displayed in Fig.\,\ref{fig:Panel3}e sums to 12.95\,mW, leaving 1.51\,mW in the remainder of the spectrum.\\

\noindent\textbf{Pulse characterization}
\\
For the measurement of Fig. \ref{fig:Panel4}d, we supplied 107.9 mA of current to the SOA, achieving a 2.33 mW on-chip average power, with 151 µW of injected power at 1550 nm. We detuned the injection from the cavity resonance to a value similar to the middle panel shown in Fig. \ref{fig:Panel3}d, in order to space the pulse pair formed by the LEO comb to a position in which we could resolve the floor value around the central feature of the autocorrelation measurement. Assuming a Gaussian pulse shape, the inferred FWHM pulse length was 1.76\,ps, which, for a pulse pair repetition rate of 17.395 GHz, yielded a peak power of 35.8\,mW, corresponding to a peak power 237 times larger than the optical injection.\\

\noindent\textbf{Simulation parameters}
\\
To closely reproduce the measured dynamics, experimentally determined parameters were used whenever available. In our system, we estimate a round-trip loss of 7.5 dB, which includes both internal losses and the output transmission, with $\theta$ = 0.5. This corresponds to $\ell \approx 1.73$ via $L_{dB}=10\log_{10}\left(\exp(-\ell)\right)$. The SOA saturation power is $P_\text{sat}=6\,$mW, and the average cavity group delay dispersion is 4280 fs$^2$. The third-order nonlinearity of the SOA is not well established; therefore, we adjusted this parameter within a reasonable range to achieve the closest agreement with experiment, which yields $\gamma=10$ (W$\cdot$m)$^{-1}$. Finally, we set the gain bandwidth of the SOA to 25 nm. We set the modulation amplitude, $M$, to 1.6 radians and the modulation frequency $\omega_m/2\pi$ to 17.395 GHz, which corresponds to twice the cavity free spectral range (8.7 GHz, i.e.,  2$\pi$ radians) and an angular modulation frequency of $\omega_m=1.749\times10^{11}$ rad/s. \\\indent
For Fig. \ref{fig:Panel2}, we numerically integrate Eq. (1). Then, while fixing both the injection power at $P_{in} =500$\,µW, and detuning at $\delta_0 = 0$, we gradually increase the gain up to 5 times above threshold while plotting the effective loss and intracavity power in Fig. \ref{fig:Panel2}a, as well as the intracavity intensity along the fast axis in Fig. \ref{fig:Panel2}b. For the comparison plots (Fig.\,\ref{fig:Panel2}c,d), we bias the gain chip at 1.4$\times$ above threshold, giving $g_0 \approx  2.42$. To model the PM-MLL, we keep all parameters identical to those in the LEO comb simulation, except for the optical injection power, which we set to zero (i.e., $P_{in} =$ 0). For the REO comb, we keep all parameters identical to those in the LEO comb simulation, except that we set the round-trip loss to 1\% and enforce critical coupling of the injection. We further amplify the REO comb output using the same $P_{sat}$ and $g_0$ as for the LEO comb.\\\indent
For the simulated spectra in Fig. \ref{fig:Panel3}d, marked by I, II, and III, we set the injection power to 160 µW and injection detuning values to 1.6 rad, 0.4 rad, and -1.5 rad, respectively. For the simulated autocorrelation trace in Fig. \ref{fig:Panel4}b, the injection power was set to 200 µW, and the cavity detuning was scanned from -1.6 to 1.6 rad.\\

\noindent\textbf{Data availability} 
\\
The datasets used in the current study are available from the corresponding author upon reasonable request. \\

\noindent\textbf{Code availability} 
\\
The code used in the current study is available from the corresponding author upon reasonable request. 

\vspace{-10pt}
\section*{Acknowledgements} 
\vspace{-10pt}
\noindent 
The device fabrication was performed at the Kavli Nanoscience Institute (KNI) at Caltech. The authors acknowledge contributions of Qiushi Guo in the design and fabrication of TFLN devices in the early stages of the project, and Cecile Jung’s assistance in the fabrication of TFLN devices. The authors gratefully acknowledge support from DARPA award D23AP00158, ARO grant no. W911NF-23-1-0048, NSF grant nos. 2408297 and 1918549, AFOSR award FA9550-23-1-0755, the Center for Sensing to Intelligence at Caltech, the Alfred P. Sloan Foundation, and NASA’s APRA program under 108350-399131.02.05.04.10 grant. N. E. acknowledges support from the Belgian American Educational Foundation (B.A.E.F.) and the European Union’s Horizon Europe research and innovation programme under the Marie {\fontencoding{T1}\selectfont Sk{\l}odowska}-Curie Grant Agreement No. 101103780.

\vspace{-10pt}
\section*{Author contributions} 
\vspace{-10pt}
\noindent B.K.G., Y.X., and N.E. performed experiments with help from R.R., D.T., A.T., R.S., and M.B. R. S. and M. S. contributed to the device fabrication. B.K.G., Y.X., N.E., and A.M.P. performed numerical simulations. B.K.G. and A.M. wrote the manuscript with input from all authors. A.M. supervised the project.

\vspace{-10pt}
\section*{Competing interests}
\vspace{-10pt}
\noindent B.K.G., Y.X., N.E., and A.M. are inventors on a provisional patent application (US patent application no. 63/969,845) related to this work. R.S. and A.M. are involved in developing photonic integrated nonlinear circuits at PINC Technologies Inc. R.S. and A.M. have an equity interest in PINC Technologies Inc. The remaining authors declare no competing interests.

\end{document}